\documentstyle[twoside,fleqn,espcrc2,epsf,epsfig]{article}

\newcommand{\AmS}{{\protect\the\textfont2
  A\kern-.1667em\lower.5ex\hbox{M}\kern-.125emS}}

\title{Dynamical Quark Effects in QCD}

\author{Stephan G\"usken\address{Physics Department,
 University of Wuppertal,D-42097 Wuppertal, Germany}}
               
\begin{document}

\begin{abstract}
We discuss latest results of lattice QCD simulations with
dynamical fermions. Special emphasis is paid to the subjects
of the static quark potential, the light hadron spectrum, $\Upsilon$ 
spectrum, and the pion-nucleon-sigma term. 
\end{abstract}

\maketitle

\section{INTRODUCTION}

The investigation of the influence of sea quarks on the
physics of strong interaction has been a field of vital interest
ever since the formulation of lattice quantum chromo dynamics.

However, the computer simulation of full QCD processes 
is a very challenging task, and it is only in these days, that 
computers and algorithms might become powerful enough to produce
results which can be connected reliably to continuum physics.

Several large scale lattice simulations of full QCD have been
started recently, unfortunately most of them being still
in the  phase of lattice calibration or program development. 
The CP-PACS, UKQCD and QCDSP projects belong to this
category. Status reports and first (preliminary) results 
can be found in \cite{Kaneko,Burkhalter}, \cite{Talevi,Irving},
\cite{Fleming}.
The bulk of -- at least semi-final -- results achieved this year
comes from the SESAM and the T$\chi$L \cite{Lippert}
collaborations. Therefore, these results
will serve as a guideline for this talk. In the first section I will
discuss the status of static potential measurements. The second section  
deals with slowly moving quarks, i.e. the NRQCD
results come into focus. The third section summarizes the SESAM/T$\chi$L
investigations on the light spectrum, and finally the lattice measurement   
of the pion-nucleon-sigma term $\sigma_N$ is reviewed in section 
four.
\footnote{ For flavour dependence of chiral quantities see\cite{Mawhinney,Luo}.}

\section{STATIC POTENTIAL}

One of the most exciting features that we expect to be a property
of full QCD, but not of quenched QCD, is that the energy string
between a quark and it's anti-quark should break
once the distance between these particles becomes large enough to 
produce a quark anti-quark pair out of the vacuum.
Clearly, one would like to verify this important quality in full
QCD lattice simulations. 
\begin{figure}[htb]
\epsfig{file=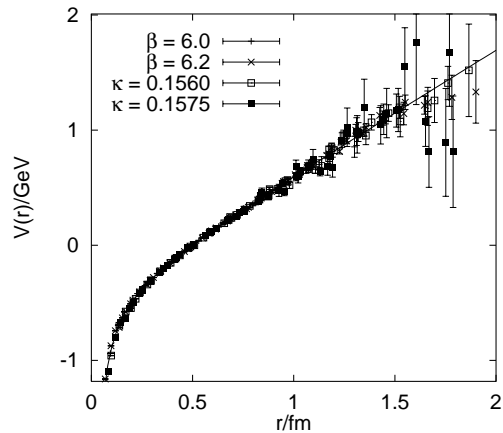,width=7.0cm}
\vskip -0.7 cm
\caption[a]{\label{Potential_wir_plot}
\it{SESAM/T$\chi$L collaboration. The rescaled static potential in full
and in quenched  QCD. The data at $\kappa=0.1575$ comes from ana\-ly\-sis
of the T$\chi$L lattice. The methods of analysis have been 
explained in detail in \cite{Potential_alt} 
.}} 
%\vskip -0.4cm
\end{figure}
One prominent quantity, which should exhibit a clear
sign of string breaking, is the potential between static quarks. The
signature that we expect is a flattening out of the potential
at a given distance $r_{break}$. Unfortunately, this has not been
observed yet in lattice QCD simulations up to distances of $r \simeq 1.5fm$
and quark masses down to approximately the strange quark mass.
The SESAM/T$\chi$L collaboration has performed a high precision measurement
of the static potential \cite{Bali} with $n_f=2$ dynamical Wilson fermions at 
$\beta =5.6$. SESAM has used a lattice size $16^3 \times 32$ at 
four different values of the sea quark mass, $\kappa_{sea} = 0.1560,0.1565,
0.1570,0.1575$, corresponding to $m_{\pi}/m_{\rho} =$ 0.839(4),
0.807\footnote{Configurations with this sea quark mass are currently
analyzed.},0.755(7),0.69(1),
with 200 statistically independent gauge configurations.
The T$\chi$L collaboration has used lattice size $24^3 \times 40$,
two sea quark masses, $\kappa_{sea} = 0.1575,0.1580$, corresponding
to $m_{\pi}/m_{\rho} \simeq 0.69,0.55$, and a statistics of presently
$\sim 80$ configurations.
The result -- at two different values of the quark mass -- is shown
in fig.\ref{Potential_wir_plot}. Obviously, there is still no clear sign
of string breaking in this range of quark masses and up to a physical distance
of $r \simeq 2.0fm$.   

\begin{figure}[htb]
\vskip -0.3cm
\epsfig{file=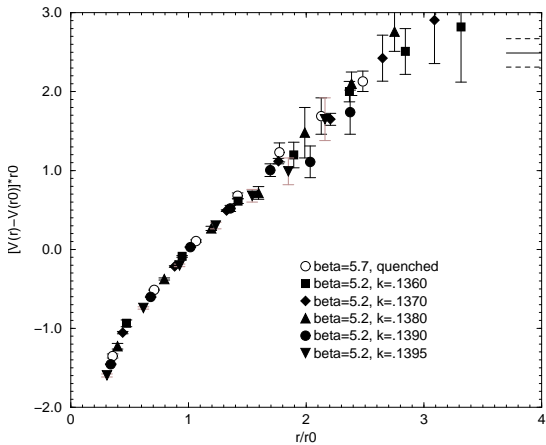,bbllx=80,bblly=54,bburx=560,bbury=640,width=6.0cm,angle=270,clip=}
\vskip -0.7 cm
\caption[a]{\label{UKQCD_pot_plot}
\it{UKQCD collaboration (preliminary). The re\-scaled sta\-tic po\-ten\-ti\-al
 in full
and in quenched ($\beta=5.7$) QCD. $r/r_0=3$ corresponds roughly to
$r\simeq 1.5 fm$. The naive expectation of string breaking is displayed
at the right edge.For details see \cite{Talevi}
.}} 
%\vskip -0.4cm
\end{figure}
The result of SESAM/T$\chi$L is supported by the (preliminary)
findings of the UKQCD collaboration\cite{Talevi}. Their result,
obtained with an improved
clover action\cite{Jansen} at $n_f=2$, $\beta=5.2$, and a lattice
volume of $12^3 \times 24$ is displayed in fig.\ref{UKQCD_pot_plot}.     
In order to  guess, at which energy $V(r_{break})$ 
the string should break, UKQCD has also calculated  the mass
of the static light meson ($M_{SL}$) 
at $\kappa_{light} =0.1390$. Naively, one expects $V(r_{break}) \simeq 2 \times
M_{SL}$. The data in fig.\ref{UKQCD_pot_plot}, however, doesn't
seem to obey this naive
expectation, although there is still a small chance that they may do so (at
$\kappa_{sea} = 0.1390$). This uncertainty can be removed soon, as
UKQCD's simulation proceeds. 

So, finally one has to tackle the question why string breaking could
not be observed yet in full QCD determinations of the static
potential. One explanation
could be that currently available sea quarks are still too heavy. The
energy necessary to produce a quark anti-quark pair out of the vacuum 
would then be available only at larger distances. This is however in
sharp contrast to the naive expectation and would therefore require 
further investigation.
 
Another explanation, which appears much more likely to the author, is
that the Wilson loop operator, which is used to calculate the potential,
does not have sufficient overlap with
the state in which the string is broken. In order to corroborate this 
scenario one would have to compare the potential energy with the 
ground state energy of an operator, which is designed to  describe
the physical situation of a
broken string. Such an operator could be that 
of a system of two static-light mesons. Its ground state would have
to be evaluated with respect to the spatial distance of the mesons. 
The crossover point between $V(r)$ and the ground state energy
of this  operator would then
determine $r_{break}$.

\begin{figure}[htb]
\epsfig{file=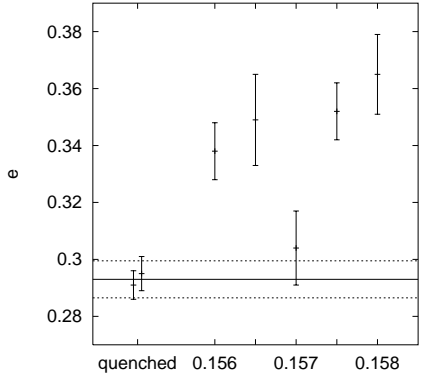,width=6.0cm}
\vskip -0.7 cm
\caption[a]{\label{Coulomb_wir_plot}
\it{SESAM/T$\chi$L collaboration. The coulomb coefficient $e$ in full
and in quenched ($\beta=6.0$) QCD. The analysis has been done analogous
to \cite{Potential_alt}
.}} 
%\vskip -0.4cm
\end{figure}
Before we turn to the discussion of NRQCD results, we would
like to emphasize that the influence of dynamical fermions can
be seen clearly in the short distance part of the static potential, i.e.
in its coulombic contribution. Fig. \ref{Coulomb_wir_plot} compares the values
of the coulomb coefficient $e$ at several sea quark masses and in
the quenched approximation. The difference between quenched and full
QCD results is statistically significant and reveals that the coulombic
part becomes stronger in full QCD. 
  
\section{$\Upsilon$ Spectrum}

The determination of heavy particle properties with lattice methods
is a notoriously difficult task, as large cutoff effects may contaminate
the results. A very promising attempt to get rid of these contaminations
is the lattice formulation of NRQCD, which, in principle, is able to
control its systematic errors. As the computational effort needed to
exploit NRQCD is small compared to standard lattice simulations, it appears
to be an ideal laboratory to study the influence of sea quarks on the 
spectrum of heavy mesons with high statistical precision.

It has been reported last year \cite{Shigemitsu}, that the
the $\Upsilon$ spectrum energy levels are shifted towards their
experimental values once sea quarks
are switched on. This behaviour had been seen in the spin independent 
as well as in the spin dependent parts of the spectrum.

\begin{figure}[htb]
%\begin{center}
\epsfig{file=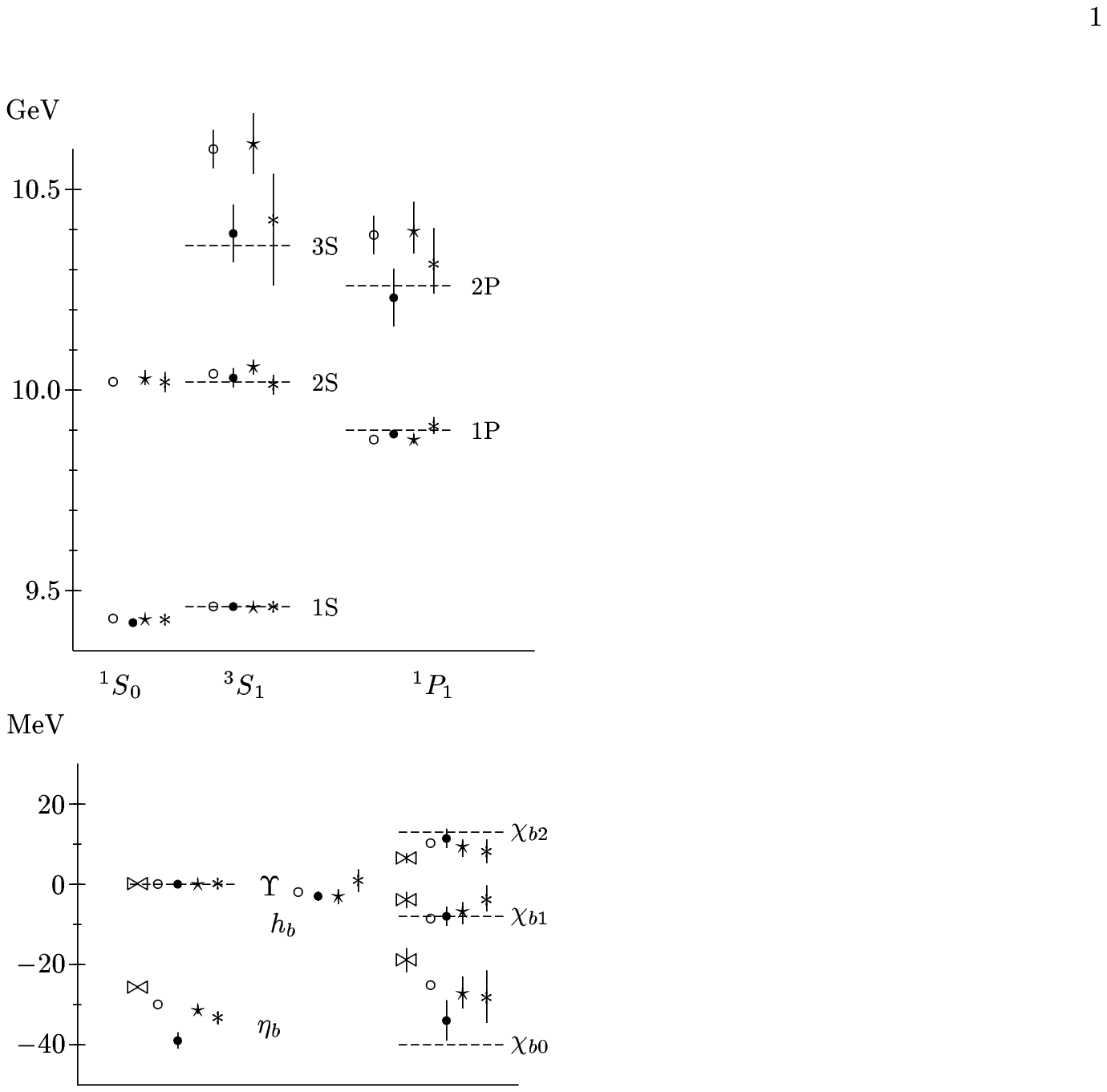,clip=,width=6.0cm}
\vskip -0.7cm
\caption[a]{\label{upsilon_plot}
\it{Spin independent (upper graph) and spin dependent parts of
 the $\Upsilon$ spectrum. Open circles:
NRQCD collab. quenched\cite{Davies}; filled circles: NRQCD collab, $n_f=2$,
KS fermions; stars: SESAM, quenched\cite{Spitz}, crosses: SESAM, $n_f=2$,
Wilson fermions, $\Join$: Manke, $n_f=0$\cite{Manke}   
.}}
%\end{center}
\end{figure}
Encouraged by this result the SESAM collaboration has now performed 
an NRQCD analysis of the $\Upsilon$ spectrum with $n_f=2$ Wilson
fermions\cite{Spitz}. With this data in hand, which we display in
fig. \ref{upsilon_plot}, the
situation now looks much less definite. Although  SESAM consolidates
the last years picture with respect to the spin independent part of the
spectrum, there is no significant effect of sea quarks in the
spin dependent part. 

In order to explain the difference between 
the results of the NRQCD collaboration\cite{Davies} and the SESAM
results, one has to remember that the definition of the NRQCD
Hamiltonian is by far not unique. The NRQCD collaboration has included
only terms up to $O(Mv^4)$ 
and used the plaquette definition of the average link for  
tadpole improvement.
SESAM went up to $O(Mv^6)$ and used the Landau gauge definition of
the average link for tadpole improvement. The effects of such
changes have been studied in detail by \cite{Spitz,Manke} 
within the quenched approximation. For the spin dependent
splittings of the $\Upsilon$ system  they find them to be of
the order $10\%-20\%$. We conclude that, using state of the art NRQCD,
unquenching effects in the spin dependent part may be
hidden by systematic uncertainties. 
 
\section{LIGHT SPECTRUM}

The accurate calculation of the light hadron spectrum, including the strange
quark sector, is still an open problem for full lattice QCD. Apart from
the lack of statistics, finite size and finite cutoff effects, one reason
is that one cannot include light and strange sea quarks 
at the same time yet.

Thus, the usual way of analyzing spectrum data has been to look at light
hadrons in a sea of light quarks only, whereas the properties of hadrons
containing strange quarks have been approximated by use of a sea
of strange quarks.  

The SESAM collaboration has now developed\footnote{We thank M.~L\"uscher
for pushing our thoughts into this direction.}
a  more general type
of analysis, which goes beyond the restriction of equal values of sea 
and valence quark mass.
With that
method one exploits a given full QCD gauge configuration -- with a
definite value of the sea quark mass $m_{sea}$ --
with respect to several valence
quark masses $m_{val}$. Guided by chiral perturbation theory one can then
parameterize hadron mass results  with respect to both
$m_{sea}$ and $m_{val}$. A {\it simultaneous} fit to all hadron data finally
allows to access arbitrary points in the ($m_{sea},m_{val}$) plane.
Of course, the validity of this ansatz has to be checked carefully with
the data. 

\begin{figure*}[htb]
\begin{center}
\parbox{15cm}{
\parbox{7cm}{\epsfig{file=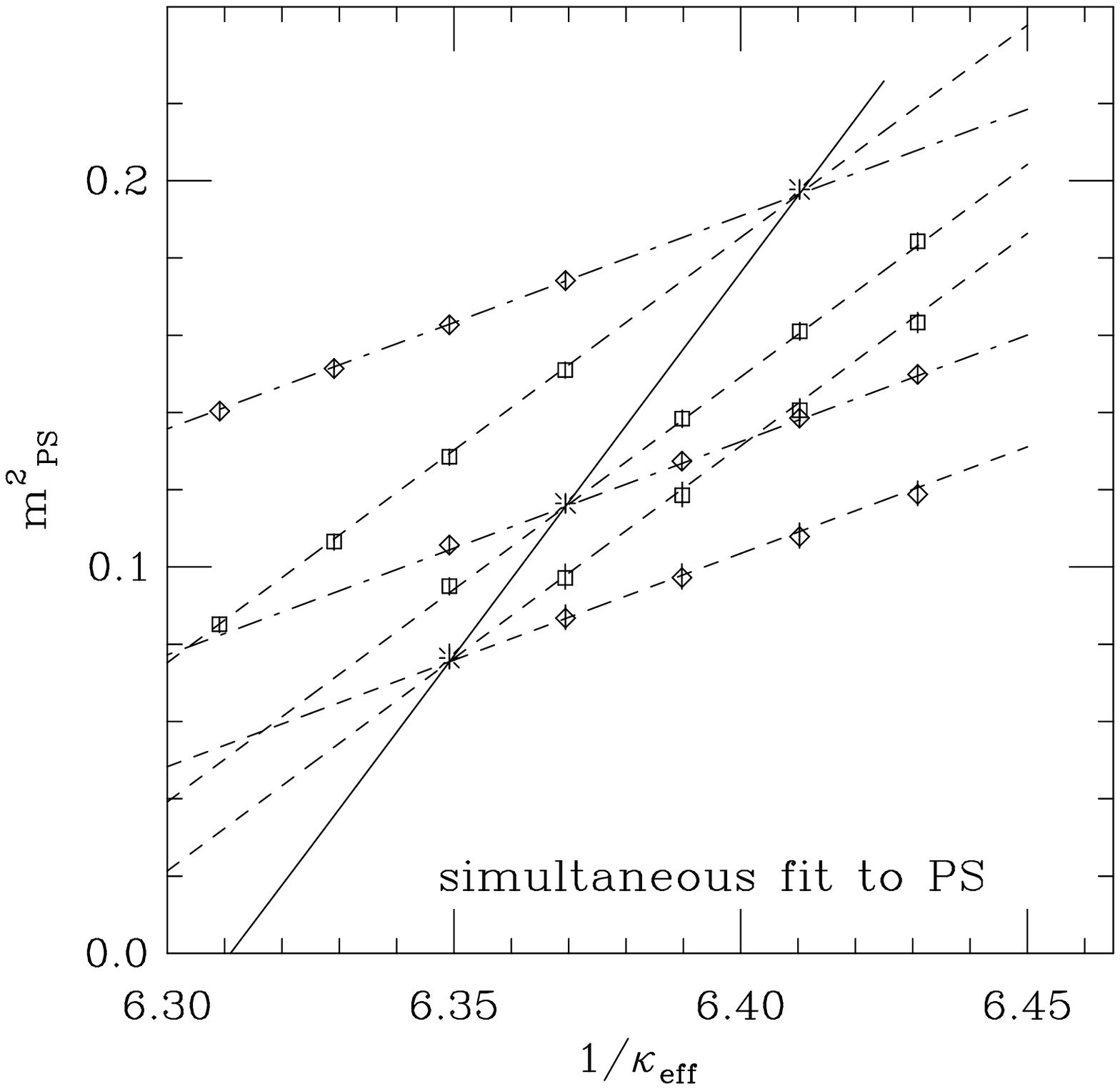,width=7.0cm}}\nolinebreak
\parbox{7cm}{\epsfig{file=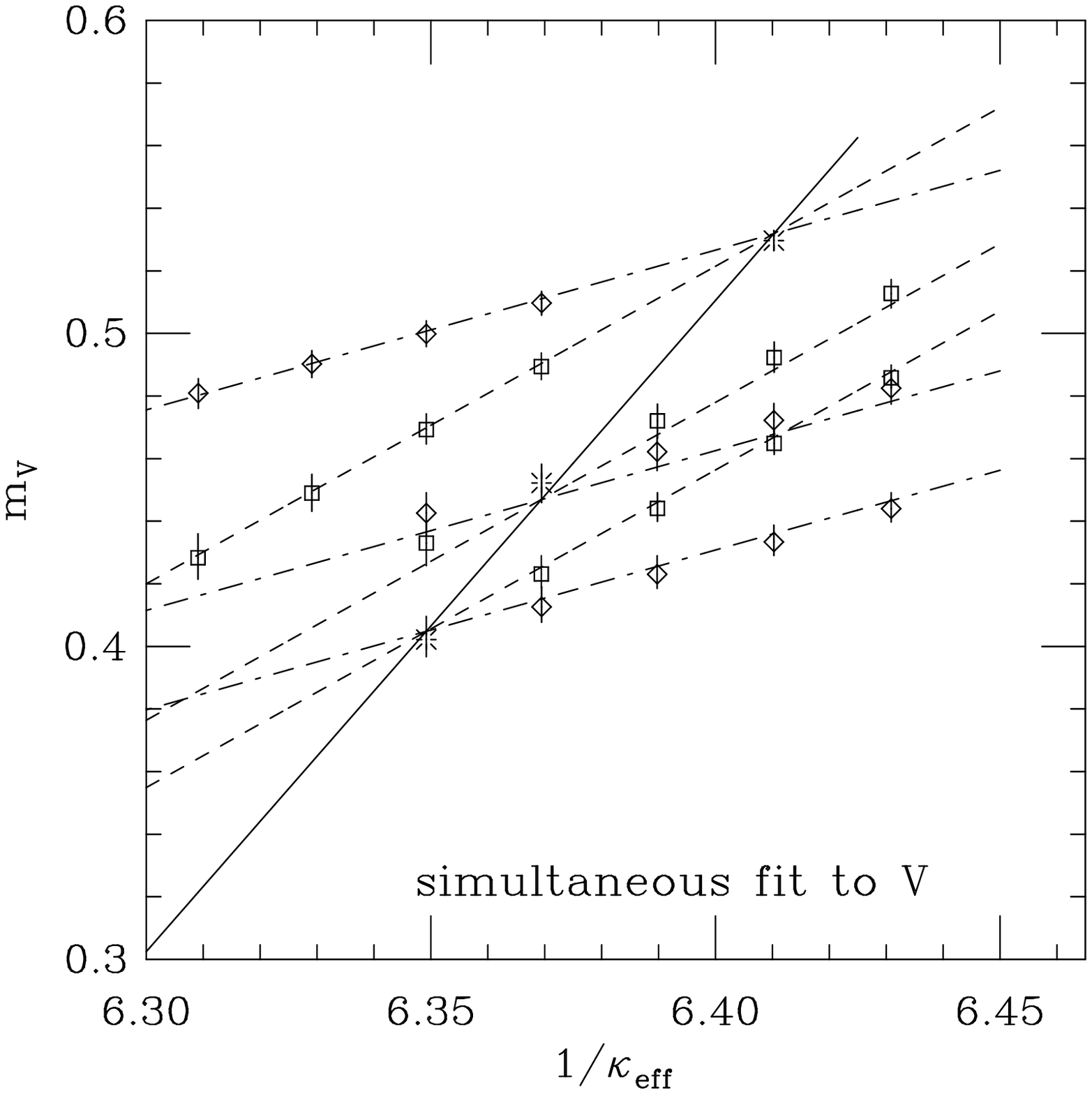,width=7.0cm}}
}
\end{center}
\vskip -1.1 cm
\caption[a]{\label{meson_vs_mq_plot}
\it{SESAM/T$\chi$L collaboration. 
Meson masses as a function of quark masses.
$m_q^{eff} = 1/2(1/\kappa_{eff} - 1/\kappa_c)$, $\kappa_{c}= 0.15846(5)$.
$m_q^{eff}$ is the average valence quark mass of the meson. Symbols:
(ss) data: $*$ and solid lines; (sv) data: $\Diamond$ and dashed-dottet lines
; (vv) data: $\Box$ and dashed lines
.}} 
%\vskip -0.4cm
\end{figure*}

SESAM has applied this method to the set of 
3 (sea quarks) $\times$ 5 (valence quarks) combinations,
with a high statistics of
200 independent gauge configurations per sea quark, using a linear
ansatz both in $m_{val}$ and $m_{sea}$. The valence quark 
masses are located in the same range as the sea quark masses (see above).
Fig. \ref{meson_vs_mq_plot} illustrates the
quality of the approach for the case of pseudo-scalar and vector mesons.
Although the data is reasonably well described  with a linear ansatz,
one cannot rule out quadratic terms in $m_q$ at this stage of the analysis.
In order to check on this SESAM is currently
collecting data at one additional value of the sea quark mass.

With the current setup -- 3 values of dynamical quarks and linear
parameterization -- SESAM finds no significant unquenching effect in the light
hadron spectrum. This is similar to the findings of the MILC collaboration
\cite{Sugar}, which uses staggered sea quarks.  

Finite size effects are a potentially dangerous source of contamination,
especially if one tries to push lattice masses towards the chiral point.
As almost nothing is known about the size of these effects in the
case of dynamical Wilson fermions\cite{Gottlieb}, the T$\chi$L collaboration
has extended the SESAM lattice to size $24^3 \times 40$. The simulation
is now being performed at the smallest SESAM sea quark mass
, $\kappa = 0.1575$,
and at an even smaller mass, $\kappa=0.1580$. With the current statistics
-- 80(200)  T$\chi$L(SESAM) configurations -- a direct comparison reveals
finite size effects of $O(5\%)$, but at this stage of the simulation it is   
clearly too early to draw definite conclusions.

\section{PION-NUCLEON-SIGMA TERM}
 
The determination of the pion-nucleon-sigma term
\begin{equation}
\sigma_N = m_l \langle P | \bar{u}u + \bar{d}d | P \rangle \; ,\;
m_l = \frac{1}{2}(m_u + m_d) 
\label{sigma_N}
\end{equation}
has a long and eventful history, reaching back into the late seventies
Although experimentally as well as theoretically difficult to determine,
it is an important quantity, measuring
chiral symmetry breaking in QCD. It is also a fascinating
quantity, as it is expected that pure vacuum fluctuations make up at least half
of its size.
Since we are looking for sea quark effects, it appears mandatory to 
study $\sigma_N$ in full lattice QCD.

The cheap way to calculate $\sigma_N$ in full lattice QCD is simply to 
determine the proton mass at several  quark masses and to apply 
the Feynman-Hellman theorem 
\begin{equation}
 m_l \langle P | \bar{u}u + \bar{d}d | P \rangle = m_l \frac{\partial m_{P}}{\partial m_l}\; . \label{F_H}
\end{equation}
However, in a statistical simulation,
the outcome of this method may be rather unprecise: suppose one would be
able to determine $m_P(m_q)$ to a precision which allows to resolve the
quadratic term of this function. Then, because of eq.\ref{F_H}, the 
quark mass dependence of $\sigma_N$ would be known only to the order linear
in $m_q$. 
Thus, in order to extract maximal information from a given set of configurations, it seems advantageous to choose the hard way, i.e. to calculate the
connected and the disconnected parts of eq.\ref{sigma_N}.
\begin{figure}[htb]
\vskip -0.2cm
\epsfig{file=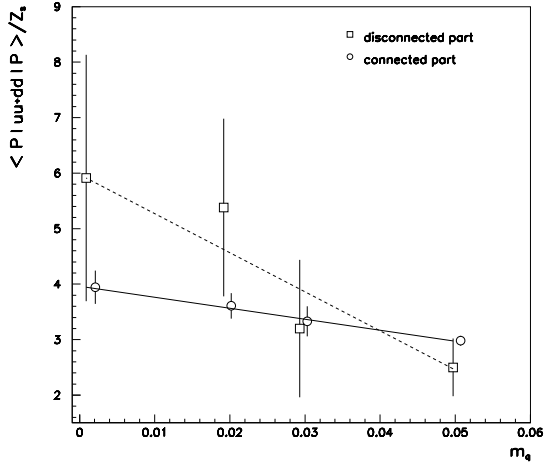,bbllx=60,bblly=190,bburx=550,bbury=620,width=7.0cm}
\vskip -0.8 cm
\caption[a]{\label{Nsigma_sesam_plot}
\it{SESAM collaboration. Connected and disconnected amplitude
$\langle P | \bar{u}u +  \bar{d}d | P \rangle$ as a function of the
quark mass. The extrapolated values are depicted at $m_q=0$
.}} 
%\vskip -0.4cm
\end{figure}

The SESAM collaboration has followed this way of analysis.
The connected part has
been determined with the standard procedures.
The disconnected contribution to $\sigma_N$ is given by the correlation
between the proton propagator and the quark loops. The latter were
calculated by application of the stochastic estimator method
with complex $Z_2$ noise. 
The statistical quality of the correlation signals
could be increased substantially by using a plateau-like sampling 
of quark loops\cite{Viehoff} instead of the usual summation
over all quark loops in space and time. 

Fig.\ref{Nsigma_sesam_plot} displays the SESAM result for the 
(unrenormalized) amplitude $\langle P | \bar{u}u + \bar{d}d |P \rangle$.
Both contributions, connected  and disconnected, increase as the 
quark mass  decreases. A linear extrapolation to the light quark mass
yields
\begin{equation}
\sigma_N = 19.65(4.39) \mbox{MeV}\;.
\label{sigval_sesam}
\end{equation}
Here we have multiplied with the
light quark mass, $m_l = 0.5(1/\kappa_l - 1/\kappa_c)$, $\kappa_l=0.158415,
\kappa_c=0.158458$, and with the inverse lattice spacing, $a^{-1}=2.33$GeV.
This value (eq.\ref{sigval_sesam}) is almost a factor
two smaller than the results obtained
from phenomenology\cite{Leutwyler} and from quenched
QCD\cite{Tsukuba,Kentucky}. We emphasize that no renormalization factor is
involved here, as $\sigma_N$ is a renormalization group invariant quantity. 
\begin{figure}[htb]
\vskip -0.2cm
\parbox{7.0cm}{
\parbox{3.5cm}{\epsfig{file=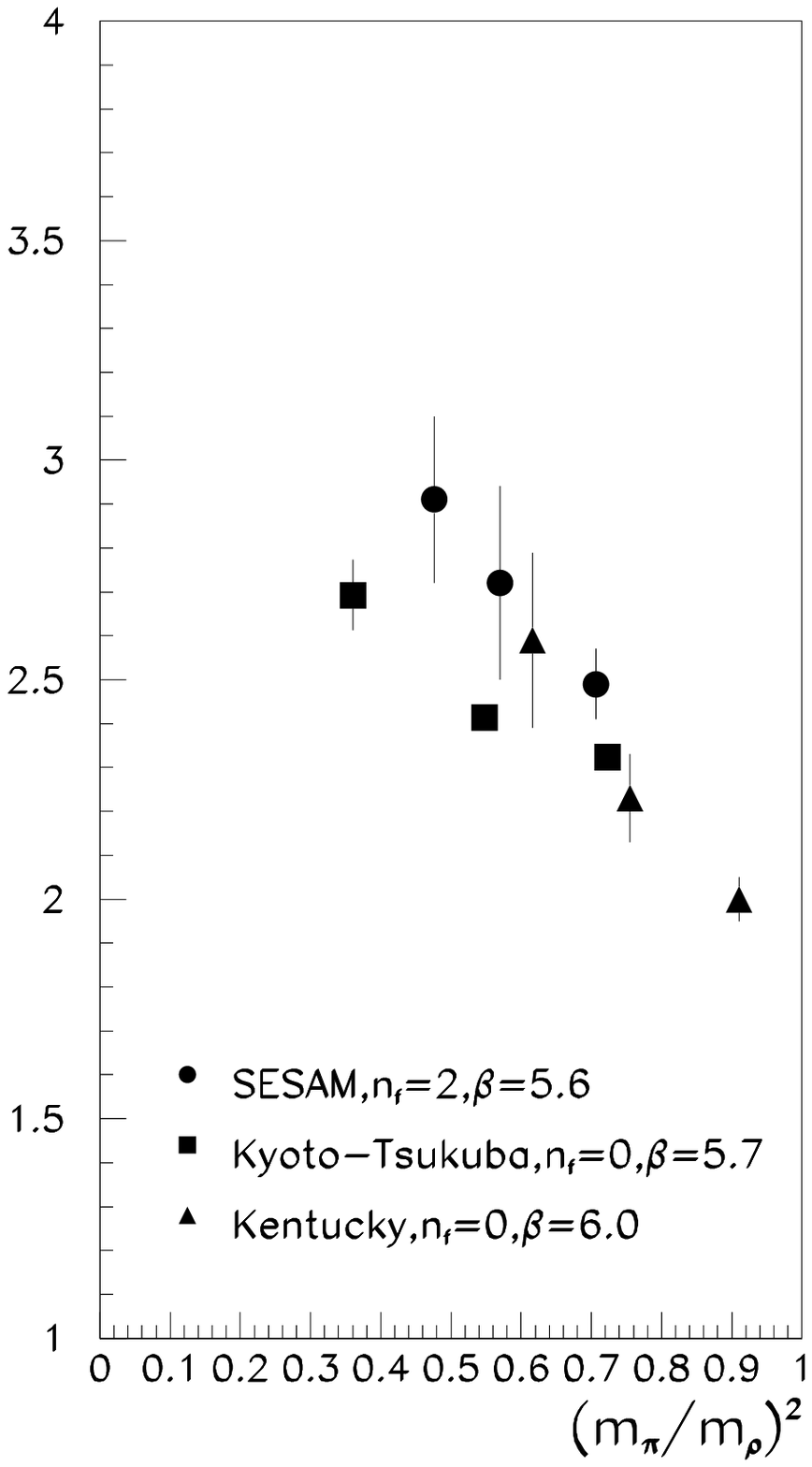,width=3.5cm}}\nolinebreak
\parbox{3.5cm}{\epsfig{file=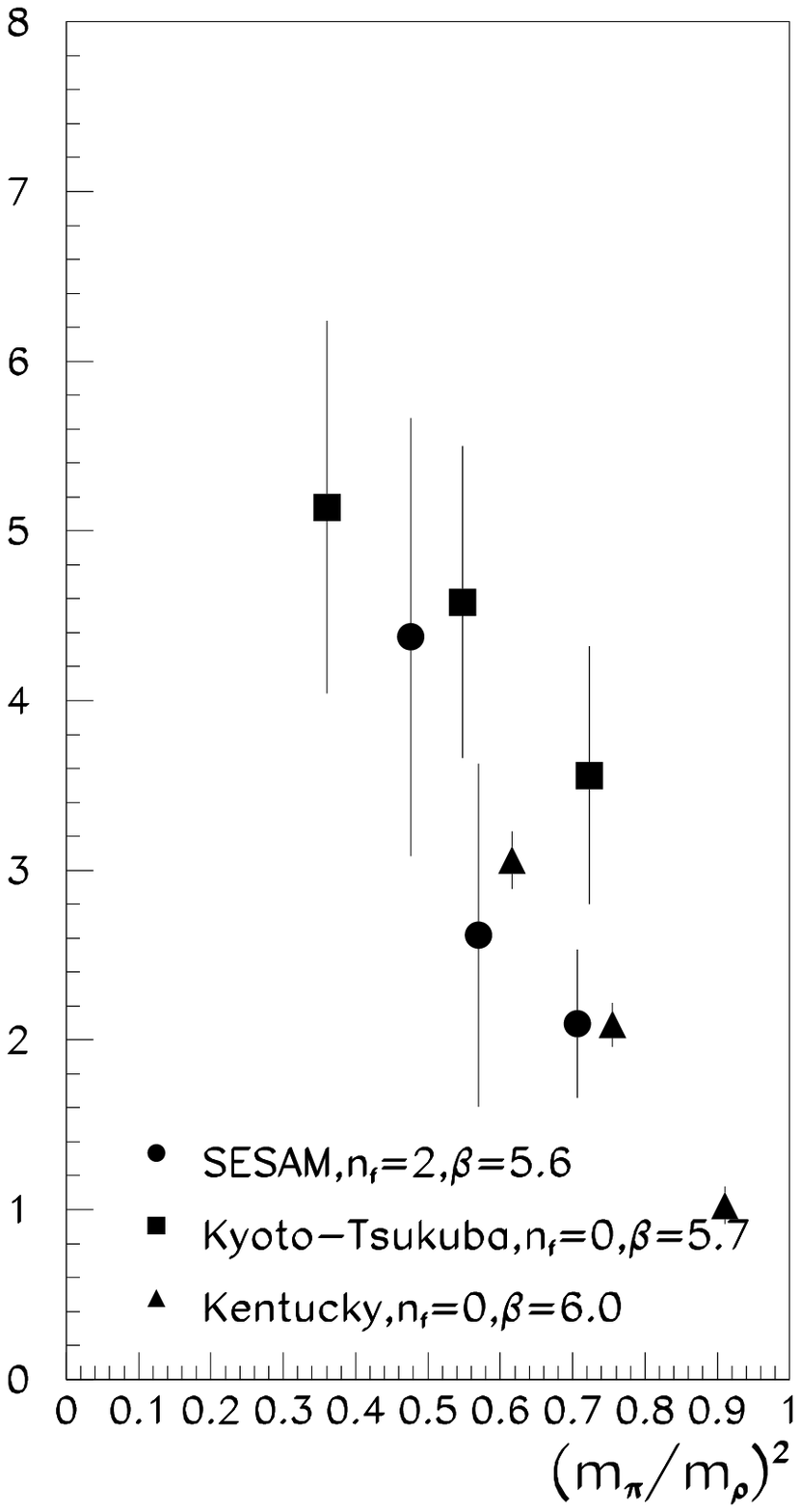,width=3.5cm}}
}
\vskip -0.8 cm
\caption[a]{\label{Nsigma_compare}
\it{Connected (left) and disconnected contributions to 
$\langle P | \bar{u}u + \bar{d}d | P\rangle$. The data is taken from
\cite{Tsukuba,Kentucky} and has been renormalized with the standard 
tadpole improved renormalization factor $Z_S$
.}} 
%\vskip -0.4cm
\end{figure}

In order to find out where this large deviation comes from, we compare in 
fig. \ref{Nsigma_compare} the SESAM results with previous quenched results. 
Obviously all amplitudes are in reasonable agreement. Thus we conclude 
that the large deviation in $\sigma_N$ is caused by the large difference
 between quenched and unquenched 
quark masses\footnote{This has been anticipated by M.~Okawa\cite{Okawa}.},
found in \cite{Gupta_old,Hoeber}.
It is an open issue\cite{Gupta_lat97}, whether present lattice techniques
are sufficient to determine $m_l$ reliably in full QCD. But
lattice calculations of $\sigma_N$ appear
to exclude already that uncertainties due to mass renormalization factors
could be responsible.
%%\vspace*{-0.3cm}
\section{CONCLUSIONS}

Sea quarks are shy objects. The only places where we could at least
see the footprints of Wilson fermions are the short
distance part of the static 
potential and the spin independent part of the $\Upsilon$ spectrum.
Obviously very accurate simulations, very refined analysis methods
and a clear understanding of the underlying physics 
are needed to observe sea quarks at work. The first promising
steps in this direction have been done recently, and we are looking
forward to the next generation of full QCD simulations, namely UKQCD,
CP-PACS and QCDSP.

\end{document}